\documentclass[conference]{IEEEtran}
\IEEEoverridecommandlockouts
\usepackage{cite}
\usepackage{amsmath,amssymb,amsfonts}
\usepackage{graphicx}
\usepackage{textcomp}
\usepackage{xcolor}
\usepackage{color}
\usepackage{wrapfig}
\usepackage{balance}

\usepackage{booktabs}
\usepackage{multirow}
\usepackage{float}
\usepackage{placeins}
\usepackage{eufrak}
 \usepackage{tikz} 
\usepackage[flushleft]{threeparttable}
 
\usepackage{framed}
\usepackage{caption}
\usepackage{subcaption}
\usepackage{graphicx,xcolor} 
\usepackage[framemethod=tikz]{mdframed}
\usepackage{algorithm}
\usepackage[noend]{algpseudocode}
\usepackage{lipsum}% http://ctan.org/pkg/lipsum
\usepackage{url}
\usepackage{tikz}
\usepackage[color=black]{background}%just for under review note

\definecolor{antiquewhite}{rgb}{0.98, 0.92, 0.84}
\definecolor{azure}{rgb}{0.94, 1.0, 1.0}

\def\BibTeX{{\rm B\kern-.05em{\sc i\kern-.025em b}\kern-.08em
    T\kern-.1667em\lower.7ex\hbox{E}\kern-.125emX}}
    
\def\HiLi{\leavevmode\rlap{\hbox to \hsize{\color{yellow!50}\leaders\hrule height .8\baselineskip depth .5ex\hfill}}}

\newcommand{\funclabel}[1]{%
  \@bsphack
  \protected@write\@auxout{}{%
    \string\newlabel{#1}{{\jayden@currentfunction}{\thepage}}%
  }%
  \@esphack
}
\usepackage{hyperref}

\hypersetup{
    colorlinks=true,
    linkcolor=blue,
    filecolor=magenta,      
    urlcolor=blue,
    pdftitle={Overleaf Example},
    pdfpagemode=FullScreen,
    }
    
\begin{document}

% \title{A High-Throughput Accelerator Design For Deformable Convolution Networks using \\High Bandwidth Memory}
% \title{A High-Throughput Acceleration Paradigm For Deformable Convolution Networks using \\High Bandwidth Memory}
\title{Accelerating Monte-Carlo Tree Search on CPU-FPGA Heterogeneous Platform}

\author{\IEEEauthorblockN{Yuan Meng\IEEEauthorrefmark{1}, Rajgopal Kannan\IEEEauthorrefmark{2}, Viktor Prasanna\IEEEauthorrefmark{1}}
\IEEEauthorblockA{\textit{University of Southern California\IEEEauthorrefmark{1}, US Army Research Lab\IEEEauthorrefmark{2}}}}
% \IEEEauthorblockA{\textit{Ming Hsieh Department of Electrical and Computer Engineering, University of Southern California \\
% Contact: \{ymeng643,prasanna\}@usc.edu}}}

\makeatletter
\patchcmd{\@maketitle}
  {\addvspace{0.5\baselineskip}\egroup}
  {\addvspace{-1\baselineskip}\egroup}
  {}
  {}
\makeatother
\maketitle

\SetBgContents{ } %used to be under review
\SetBgScale{1}
\SetBgAngle{0}
\SetBgPosition{current page.north east}
\SetBgHshift{-2cm}
\SetBgVshift{-1cm}

\begin{abstract}
Monte Carlo Tree Search (MCTS) methods have achieved great success in many Artificial Intelligence (AI) benchmarks. 
% MCTS manages a decision tree as the policy to guide an AI agent towards an optimal decision.
% The tree is built and updated using statistics collected in simulations.
% Parallel MCTS algorithms have been developed 
% on multi-core CPUs 
% to enable independent, parallel simulations. 
The in-tree operations become a critical performance bottleneck in realizing parallel MCTS on CPUs.
% Existing 
In this work, we develop a scalable CPU-FPGA system for Tree-Parallel MCTS. 
We propose a novel decomposition and mapping of MCTS data structure and computation onto CPU and FPGA to reduce communication and coordination.
High scalability of our system is achieved by 
encapsulating in-tree operations in an SRAM-based FPGA accelerator.
% developing an SRAM-based accelerator for in-tree operations.
% to address the bottleneck in existing multi-core CPU implementations. 
% To overcome the challenge of degraded performance from increasing number of parallel workers due to intrinsic dependencies and low-arithmetic intensity, 
% To address the challenges of 
To lower the high data access latency and inter-worker synchronization overheads,
we develop several hardware optimizations. 
% These include pipelining, tree partitioning, and quantized number representation for general MCTS policies. 
% Furthermore, to ensure the generality of the CPU-FPGA system, we provide a PyOpenCL-based programming API that allows easy customization and integration of the FPGA accelerator to interface with any existing software simulation environment using high-level libraries (e.g. OpenAI-gym, Pytorch). 
We show that by using our accelerator, we obtain up to $35\times$ speedup for in-tree operations, and $3\times$ higher overall system throughput. Our CPU-FPGA system also achieves superior scalability wrt number of parallel workers than state-of-the-art parallel MCTS implementations on CPU.
% \footnotetext{This work is supported by NSF under grant No. CNS-2009057.}
\end{abstract}
\begin{IEEEkeywords}
% component, formatting, style, styling, insert
Artificial Intelligence, MCTS, FPGA
\end{IEEEkeywords}

\section{Introduction}
Monte Carlo Tree Search (MCTS) is a planning algorithm that combines a best-first search tree with random simulations in order to find optimal decisions in a system. MCTS methods have shown a massive potential due to their success in situations where deterministic algorithms fail to deliver optimal solution in a reasonable amount of time, especially when applied in domains with extremely vast search space. For example, it is applied in many Artificial Intelligence (AI) benchmarks such as Constraint Satisfaction Problems \cite{satomi2011real}, Computer Games \cite{alphazero}, and Neural-Architecture Search \cite{wang2020neural}.

MCTS manages a tree as the policy to guide an agent towards an optimal decision. 
The tree is dynamically constructed by collecting statistics through simulations. These statistics are used to update the tree. 
% In sequential implementation of traditional MCTS methods, the bottleneck lies in the large number of complete rollout simulations performed to evaluate the quality of problem states represented by tree nodes.
% in order to achieve high rewards. 
% Therefore, existing acceleration of MCTS in the past decade focus on . For instance, recent success of AlphaZero used deep learning to replace complete stimulation rollouts with neural network inference that approximates a utility 
% Recent works have shown that rollout simulations can be easily parallelized without impacting algorithmic performance \cite{Liu2020Watch}. 
Recent advances in MCTS parallelization have shown that using multiple workers, simulations can be embarrassingly parallelized without increasing the total simulation cost to achieve the same domain performance as sequential MCTS \cite{wu-uct}. Additionally, deep neural network (DNN) has been used to replace time-consuming simulation with DNN inference \cite{alphazero}.
% that approximates utility values of states represented by tree nodes
% , further enabled more room for parallelization and thus lower latency for simulations (e.g. AlphaZero, etc).
% These advances suggest that 
The bottleneck for large-scale parallel MCTS system shifts from simulations to in-tree operations.
However, it is challenging to efficiently accelerate in-tree operations on multi-core CPUs because (1) intrinsic dependencies between workers operating on a shared tree lead to large thread-level synchronization overheads,
% in shared memory programming, 
and (2) in-tree operations have low-arithmetic intensity and incur high-latency memory access overhead.

In this work, we define the first  CPU-FPGA system for general parallel MCTS. 
% The system is composed of a hardware accelerator for in-tree operations using High-Level Synthesis (HLS), and an interface that bridges CPU-FPGA with minimized communication traffic based on our system mapping. 
Unlike existing FPGA implementations that focus on accelerating application-specific simulations \cite{jahanshahi2013blokus,qasemi2014highly}, our system's generality is in the flexibility of the accelerator to interface with the software simulations for various applications on the CPU.
% =================TODO: keep the following 1 sentence?========================
% The system is composed of a hardware accelerator for in-tree operations using High-Level Synthesis (HLS), and an interface that bridges the communication between CPU and FPGA. 
% our system provides a unified FPGA accelerator that can be customized for various applications at compile time.
% that is both generalizable to various in-tree traversal heuristics 
% and can be interfaced with any application benchmarks using software simulators.
The major contributions are:

% the first accelerator that preerves the generality of simulation CPU-FPGA heterogeneous platform.
% The accelerator is described in a parameterized template that can be customized for specific applications at compile time, and can be interfaced with any simulation software on CPU at run-time.

\begin{itemize}
    % \item We map the compute components of MCTS (simulations and in-tree operations) onto a CPU-FPGA heterogeneous platform, and define the interface between them.
    % \item We design a MCTS acceleration system on CPU-FPGA heterogeneous platform. It is composed of a hardware accelerator for in-tree operations using High-Level Synthesis (HLS), and an interface that bridges CPU-FPGA.
    % with minimized communication traffic based on our system mapping.  
    % generality refers to the portability to various applications.
    % and define the interface between them.
    % \item To reduce the PCIe data traffic and memory burden on FPGA, we decompose the tree in MCTS into Simulation Hash (ST) and Upper Confidence-bounded Tree (UCT), and propose an efficient implementation of the ST with trivial ($O(1)$) additional operational complexity on CPU.
    \item We propose a novel decomposition of the MCTS decision tree into two separate data structures, State Table (ST) and Upper Confidence-bounded Tree (UCT).
    We show that this reduces the PCIe data traffic and memory requirement on the FPGA.
    % with additional constant time overhead on the CPU.
    % a trivial cost of constant additional time complexity on CPU.
    % To address the bottleneck in the system, w
    \item We map the UCT onto the FPGA and develop an SRAM-based accelerator for in-tree operations  with 
    % several hardware and data format 
    optimizations specialized for MCTS, including:
    \begin{enumerate}
        % \item Adjacency Array layout enables maximal streaming accesses in all in-tree operations (selection, back-propagation and tree-shuffling).
        \item A hardware design that exploits both data- and pipeline-parallelism for in-tree operations.
        \item Tree partitioning and mapping to SRAM banks to eliminate bank conflicts in in-tree operations.
        % \item Fixed-point number representation tailored for MCTS policy to reduce pipeline stalls. 
        % \item A hierarchical comparison look-up table design to minimize the pipeline initial interval in MCTS in-tree operations.
        % compute cycles and fabric consumption.
    \end{enumerate}
    \item We provide an efficient CPU-FPGA interface for coordinating communication with the accelerator and synchronizations among the CPU threads in Tree-Parallel MCTS.
    % \item 
    % \item We develop a software API that generates accelerator for in-tree operations given MCTS hyper-parameters and specified number of parallel workers at compile-time, and directly ports to the python simulation program on CPU at run-time.
    \item We replace the CPU process for in-tree operations with the FPGA accelerator in two state-of-the-art parallel MCTS implementations.
    % (traditional simulation-based and DNN-based). 
    This leads to up to 35x speedup for in-tree operations, and our system achieves up to 3x higher throughput than the CPU-only implementation.
    % scalability and 
    %  of the system by .
    
\end{itemize}
\section{BACKGROUND}
% 0.5 - 1 page
\label{sec:background}
\subsection{Sequential Monte-Carlo Tree Search}
\begin{figure}
    \centering
    \includegraphics[width=0.98\linewidth]{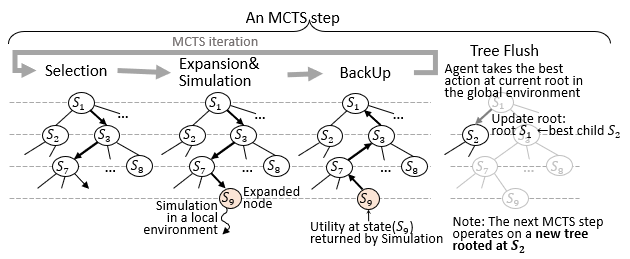}
    \caption{MCTS procedure. An MCTS iteration is defined as one round of Selection-Expansion-Simulation-BackUp by one (or multiple) worker(s). An MCTS step is the entire process that after the number of tree nodes reaches a limit $X$, the agent takes one step in the global environment and flushes the tree.}
    % after $X$ MCTS iterations by the worker(s) 
    \label{fig:background}
\end{figure}

% 0.5 page
\label{sec:seq_mct}
Monte Carlo Tree Search (MCTS) is a model-based Reinforcement Learning algorithm performed by an agent that explores a global environment, and plans the best action at each time step by constructing a tree using one or multiple workers \cite{browne2012survey}. The tree is built using simulations on copies of the global environment called local environments. 
% The tree can be constructed using one(multiple) worker(s) that simulate(s) in one(multiple) local environments. 
In the search tree, each node $s$ represents a visited state, its adjacent edge
$(s,\hat{s})$ denotes an action taken at that state, and its child node $\hat{s}$ denotes the state it transits to after taking the action. 
Fig. \ref{fig:background} shows the phases conducted by a worker. The \textbf{Selection} phase traverses the existing search tree by iteratively choosing one child node at each tree level according to Eq. \ref{eq:uct} \cite{kocsis2006bandit}
% an in-tree policy until reaching a leaf node. 
(a leaf node is a node whose child nodes $\hat{s}$ have not been all expanded):
% The most widely used 
% The in-tree policy is the Upper Confidence-Bound (UCB) for Trees \cite{kocsis2006bandit}:
\begin{equation}
\label{eq:uct}
\small
s\leftarrow \underset{\hat{s} \in \text{Children}(s)}{\arg \max }\left\{uct(s,\hat{s})\right\} = \underset{\hat{s}}{\arg \max }\left\{V_{\hat{s}}+\beta \sqrt{\frac{\ln N_{s}}{N_{\hat{s}}}}\right\}
% ; s\leftarrow s'
\end{equation}
where $V_{\hat{s}}$ is the average expected utility value (i.e., reward) that can be received through $\hat{s}$, and $N_s$ ($N_{\hat{s}}$) denotes the number of times the nodes $s$ ($\hat{s}$) has been visited. $uct(s,\hat{s})$ is the weight of the edge ($s,\hat{s}$). $\beta$ is a parameter controlling the tradeoff between exploitation (first term) and exploration (second term). 
% Each Selection initiates a simulation request.
% Suppose the leaf node returned is
% The \textbf{Expansion and Simulation} phases 
% first insert a new (un-expanded) node under the selected leaf node $s$, and then run the local environment simulation from $s$ with a simulation policy. Note that this can be done either by running a simulation software that returns the reward upon termination \cite{browne2012survey}, or by feeding the state into a DNN whose inference returns the expected $V_{s^{\prime}}$ (e.g. AlphaZero \cite{alphazero}). 
% run a local environment simulation from the selected leaf node $s$ with a simulation (random) policy, insert a new (un-expanded) node $s'$ under $s$ based on the immediate new action acted on $s$, and returns a reward $V_{s'}$ when the simulation is terminated. Note the simulation is either a software that runs until termination \cite{browne2012survey}, or a DNN whose inference approximates the expected $V_{s^{\prime}}$ (e.g. AlphaZero \cite{alphazero}).
When leaf node $s$ is selected,
the \textbf{Expansion} phase chooses a new (un-expanded) edge $(s,s')$, runs local environment 1-step simulation from the state of $s$ to reach a new state, and inserts a newly expanded child node, $s'$ that represents the new state.
The \textbf{Simulation} phase runs local environment simulation from $s'$,
% with a simulation policy,
and returns a reward $V$. Note that this can be done either using a simulation software that runs until termination \cite{browne2012survey}, or via  a DNN inference that approximates the expected $V$ \cite{alphazero,gomoku}.
Finally, the \textbf{BackUp} phase uses the received $V$ to update the $V_{\hat{s}}$ term of $uct$ for all the traversed edges during Selection.
In practice, the above phases are repeated until a resource budget $X$ (the maximum number of nodes maintained by the tree) is reached \cite{browne2012survey}. Afterwards, the agent takes a step in the global environment by choosing the best action (based on Eq. \ref{eq:uct}) at the root, and the best child becomes the new root while the rest of the tree are flushed (i.e., no longer accessed by in-tree operations in future MCTS steps). We call this a Tree Flush (Fig. \ref{fig:background}).
% It takes place only once every MCTS step, and 

\vspace{-5pt}
\subsection{Tree-Parallel MCTS}
\label{sec:pmcts}
There are many approaches to parallelize MCTS (Sec. \ref{sec:related_parallel}). The most popular method among these is the Tree-Parallel MCTS, because it can scale to larger number of workers without increasing the total simulation cost for achieving certain reward \cite{wu-uct}. In Tree-Parallel MCTS, after a worker traverses a certain path during Selection, a virtual loss \textit{VL} is subtracted from $uct$ of selected edges to lower their weights, thus encouraging other workers to take different paths. It also creates dependencies between workers during the Selection phase. \textit{VL} is recovered later in the BackUp phase. Note that \textit{VL} can either be a pre-defined constant value \cite{treeP}, or a number tracking visit counts of child nodes \cite{wu-uct}.

In Tree-Parallel MCTS, we refer to in-tree operations as all the operations that access the tree in Selection, Expansion and BackUp phases. They are defined in Algorithm \ref{alg:in-tree}.
% This includes the edge updates in BackUp, node sampling and edge updates in Selection, and node insertion in Expansion.}
% Upon each completion of Selection and Expansion of a worker, a simulation request is made.
\begin{algorithm}
\caption{\textbf{In-tree Operations} for a single worker in Tree-Parallel MCTS}\label{alg:in-tree}
\begin{algorithmic}[1]
\small
\Procedure{Selection}{}
    \State $s\leftarrow$ root, $i=1$;
    \While{$s$ is not leaf and tree level $i\in[1,D]$} 
    \State 
    % \colorbox{antiquewhite}{\strut 
    $\hat{s}\leftarrow $  RHS of Eq. \ref{eq:uct}; \Comment{\textcolor{blue}{Read $uct$ from tree level $i$}}
    % }
    \State 
    % \colorbox{antiquewhite}{\strut 
    $uct(s,\hat{s})-=$\textit{VL}; \Comment{\textcolor{blue}{Apply $VL$: Write to tree level $i$}}
    % }
    \State 
    % \colorbox{antiquewhite}{\strut 
    $E_t[i]\leftarrow (s,\hat{s}); $ $s\leftarrow \hat{s},i++;$  
    % }
    % \State 
    % \colorbox{antiquewhite}{\strut 
    % $s\leftarrow s',i++;$  
    % \Comment{\textcolor{blue}{Proceed to the next level}}
    % }
    \EndWhile
    \Return $E_t,s$ 
\EndProcedure
\Procedure{Expansion}{$s$}
    \State send state($s$) to Simulation process;
    % \If{$s$ has unexpanded child $s'$}  \Comment{\textcolor{blue}{Read from tree}}
    \State send unexpanded action($s,s'$) to Simulation process;
    \State receive state($s'$) from Simulation process;
    \State tree.InsertNode($s'$), InsertEdge($s,s'$); 
    % \Comment{\textcolor{blue}{Write to tree}}
    % \EndIf
\EndProcedure
\Procedure{BackUp}{$E_t$}
    \State receive reward $V$ from Simulation process;
    \For{ tree level $i\in[1,D]$} 
    \State 
    % \colorbox{azure}{\strut 
    tree.UpdateEdge($V$,$VL$,$uct(E_t[i])$)  \Comment{\textcolor{blue}{Update tree nodes}}
    % }
    \EndFor
\EndProcedure
\end{algorithmic}
\scriptsize
Note: $D$ is the tree height limit. 
$E_t$ and $s$ denote the traversed edges and selected leaf node in the Selection phase. $s'$ denotes the node to be expanded. state($s$) denotes the application-specific environment state represented by node $s$.
\textit{VL} is the virtual loss. 
Lines with highlighted comments are the critical regions that create Read-After-Write (RAW) or Write-After-Write (WAW) dependencies between pair of workers.
% \note{StateMap is a hash table that maps node index to environment states}
\end{algorithm}
% 0.5 page
\vspace{-5pt}
\subsection{Motivation and Objective}
A typical Tree-Parallel MCTS system is composed of a master process coordinating the in-tree operations for all workers,
% in Selection and BackUp, 
and multiple simulation processes (threads) responsible for independent Simulations \cite{wu-uct}. 
% The \textbf{system throughput} is defined by the number of simulation requests processed per unit time.
Although Tree-Parallel MCTS enables high-throughput simulation that is scalable to larger number of workers, the in-tree operations (especially Selection)
become the bottleneck.  
% as the number of parallel workers and tree depth scale up, . 
We profile the execution time breakdown of two implementations: a) parallel MCTS for Atari games \cite{wu-uct} using OpenAI-gym simulation software, and b) parallel implementation of MCTS for Gomoku game using DNN-based simulation \cite{gomoku} on a CPU. 
% As shown in Fig. \ref{fig:breakdown}, 
We observed that the Selection and BackUp total latency in an MCTS iteration grows from 10\% (8 workers) to 40\% (128 workers) for a), and 38\% (2 workers) to 70\% (32 workers) for b).
% in an MCTS step 
% \begin{figure}[h]
%     \centering
%     \subfigure[]{\includegraphics[width=0.24\textwidth]{figs/atari_bg.jpg}} 
%     \subfigure[]{\includegraphics[width=0.24\textwidth]{figs/gomoku_bg.jpg}}
%     \caption{(a),(b): Execution Time Breakdown. 
%     }
%     \label{fig:breakdown}
% \end{figure}
Such bottleneck is hard to eliminate on a multi-core CPU even with one thread per worker for Selection, because the expensive thread-level synchronizations are needed to resolve dependencies between workers due to race conditions. As shown in Algorithm \ref{alg:in-tree}, lines 4-5 (line 15) is the RAW (WAW) critical region that protect the shared tree to be only accessed by a single thread at any time. 
% This results in large latency overheads proportional to the number of workers.
This essentially serializes Selection by different workers by a synchronization interval, $T_{sync}$.
% $T_{sync}$ on CPU involves 2 threads sequentially reading and writing the same (part of) shared tree stored in the last-level cache or DRAM.
$T_{sync}$ on CPU involves the sequential process where the same 
% (e.g., level of) 
shared tree in DRAM is read by a thread after written by another thread.
This incurs high memory access latency (i.e., $T_{sync}\geq 2\times T_{mem}$, where $T_{mem}$ is the tree data access latency. Typical $T_{mem}>100$ CPU cycles \cite{SkyLake}). 
As a result, the system throughput is constrained by the latency of the Selection phase.
% which does not increase with increasing $p$ since it is bounded by large $T_{sync}$.

Motivated by the above observations, in this work, we aim to alleviate the bottleneck in Tree-Parallel MCTS system by developing a SRAM-based accelerator. 
% while preserving the dependencies in Tree-Parallel MCTS.
% that takes advantage from near-memory computing.
We (1) utilize the FPGA SRAM capacity and bandwidth to minimize $T_{mem}$ to a single FPGA cycle; and (2) exploit deep pipelining that minimizes the interval $T_{sync}$, such that the system throughput on a typical CPU-FPGA platform can be scaled to more workers without being bounded by the Selection phase.
% the end-to-end latency of in-tree operations does not increase significantly with increasing number of workers $p$. 

% By eliminating the memory access and synchronization overheads that frequently occur-ed on CPU. 
% To address the above two challenges on CPU, we aim to (1) use stall-free hardware pipelining to alleviate expensive software synchronization overhead, and (2) use abundant on-chip SRAM to store the UCT graph and specialized memory layout to achieve single-cycle access in each edge traversal step, and eliminate the memory access overhead that otherwise frequently occur-ed on CPU.

% \input{4_algorithm}
\vspace{-2mm}
% 1 page
\section{System Design}
\label{sec:system}
% \subsection{System Overview and Task Mapping}
% figure: WU-UCT and FPGA thread,
% 0.25$\sim$0.5 page
% An MCTS implementation can be viewed as an iterative process where parallel simulation processes and in-tree operators update a shared decision tree. 
% To map the memory and compute components of general MCTS to CPU-FPGA system, 
We consider a CPU-FPGA heterogeneous platform interconnected by PCIe interface.
We map the Simulation phase (and the 1-step simulation for the Expansion phase) of $p$ workers onto $p$ CPU Simulation threads.
% to take advantage of multi-core parallelism and to keep the system generalizable to various simulation tools. 
We map the in-tree operations by all the workers onto the FPGA.
% =================TODO: DELETE the following sentence!====================
% The accelerator is encapsulated as a monitor \cite{hoare1974monitors}.
% As shown in Fig. \ref{fig:background}, 

\subsection{Memory component decomposition and mapping} The tree in MCTS stores the nodes, weighted edges 
% (edge weights are $uct$ values in Eq. \ref{eq:uct}) 
and application-specific environment states represented by all the leaf nodes.
% whose edges (actions) have not been completely expanded. 
The size of the tree is thus $O(X\gamma)$, where $X$ is 
% a pre-defined algorithm paramter (i.e. 
the number of tree nodes, and $\gamma$ is the memory requirement for storing a simulation environment state. Simply storing the entire tree on the FPGA device results in $O(p\gamma)$ PCIe data traffic during the Expansion phase, as shown in Algorithm \ref{alg:in-tree} lines 8,10.
To alleviate the memory burden on the FPGA accelerator and reduce the PCIe traffic, we decompose the tree into two memory components: Upper Confidence-bounded Tree \textbf{(UCT)} and State Table \textbf{(ST)}. The UCT maintains the node and edge information in the tree except the environment states. 
The ST is implemented as a table with $X$ entries, where the index of each entry is a unique node index maintained in the UCT,
and the value is an application-specific environment state represented by that node.
The UCT is stored on the FPGA and the ST is stored in the CPU DRAM. 
% (no two nodes have the same index, therefore no hash collision is possible).
\begin{figure}[h]
    \centering
    \includegraphics[width=0.97\linewidth]{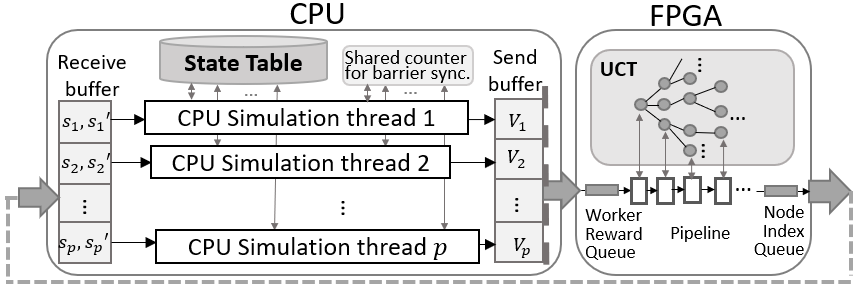}
    \caption{System Diagram and Interface}
    % , $n$ is the number of data parallel pipelines, where $p$ is the total number of workers and $A$ is the maximum fanout of the tree.}
    \label{fig:sys}
\end{figure}
Accordingly, the Expansion phase of a worker is decomposed into two parts: Node Insertion to UCT and State Insertion to ST. 
All the in-tree operations operating on the UCT in Selection,  Node Insertion and BackUp are accelerated by the FPGA.
The interactions with respect to ST (CPU) and UCT (FPGA) are further explained in Sec. \ref{sec:bsp}.
% phase is also decoupled into two parts: the node insertion on FPGA, and the ST insertion on CPU, as shown in Fig. \ref{fig:sys}. This reduce the data traffic before the Simulation phase to $O(p)$, since the FPGA only need to send $p$ selected/expanded node indices, and CPU threads use these to sample the states locally from the ST (ST Sampling). Similarly, expanded environment states do not need to be sent back to the FPGA as they are simply appended to the ST (ST Insertion).
% based on state transitions.
% reduce the data traffic before the Simulation phase to $O(p)$ by avoiding round-trip traffic from communicating environment states after the Selection phase. Specifically, after Selection and UCT node insertion on FPGA, only $p$ selected leaf nodes $s$ and expanded nodes $s'$ need to be sent from the FPGA to the CPU. For the Expansion, the CPU Simulation threads sample the environment states $e_{s}$ from the ST in parallel using $s'$, generate a new environment state $e_{s'}$ from and append new key-value pairs ($s',e_{s'}$) to the ST.
This decomposition reduces the PCIe traffic to $O(p)$ by eliminating the need to communicate large simulation states between CPU and FPGA during Expansion (only node indices need to be transferred in lines 8, 10 of Algorithm \ref{alg:in-tree}).
Additionally, the memory requirement on the FPGA is reduced since the UCT only requires $O(X)$ memory. For typical MCTS game benchmarks (e.g. Atari games and board game benchmarks \cite{wu-uct,gomoku}), $X$ ranges from $1K-100K$. This means that the entire UCT can typically be stored on the FPGA SRAM with tens of megabytes capacity.
% since each node and edge only stores 2 words data, and the complete UCT requires up to $\sim 3.3 MB$ storage, which is smaller than the SRAM capability on typical cloud FPGAs.

% Note to self: remember to clarify that with tree flush and $T_s$ constraint, for game benchmarks we consider, maintained tree size can typically fit on on-chip in FPGA SRAM.
% whose keys are the destination node indices of the expanded edge. 
\begin{figure*}
    \centering
    \includegraphics[width=0.95\linewidth]{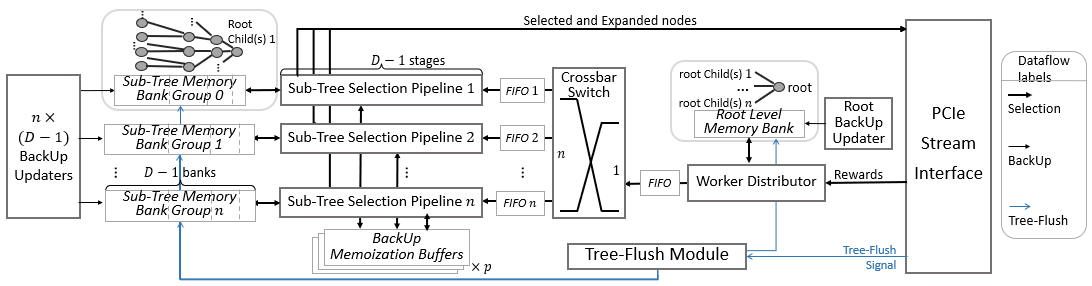}
    \caption{In-tree Operations Accelerator Overview. $D$ is the tree height limit.}
    % , $n$ is the number of data parallel pipelines, where $p$ is the total number of workers and $A$ is the maximum fanout of the tree.}
    \label{fig:arch}
\end{figure*}
% \begin{algorithm}
% \caption{FPGA kernel operations}\label{alg:fpga}
% \begin{algorithmic}[1]
% \small
% \Procedure{BackUp:}{}
%     \State receive $V$ from CPU sender buffer; \Comment{\textcolor{blue}{PCIe O($p$)}}
%     \State Do Algorithm \ref{alg:in-tree} line 14-15 for $p$ workers;
% \EndProcedure
% \Procedure{Selection:}{}
%     \State Do Algorithm \ref{alg:in-tree} line 1-7 for $p$ workers;
%     % \Return $E_t$,$idx_s$
% \EndProcedure
% \Procedure{Node Insertion}{}
%     % \If{$s$ has unexpanded child $s'$}
%     \State UCT.InsertNode($s'_{1...p}$), InsertEdge($s_{1...p},s'_{1...p}$); 
%     % \Comment{\textcolor{blue}{Node Insertion}}
%     \State send $s_{1...p},s'_{1...p}$ to CPU receiver buffer; \Comment{\textcolor{blue}{PCIe O($p$)}}
%     % \State send action($s,s'$) to Simulation process; \Comment{\textcolor{blue}{PCIe O($p$)}}
%     % \EndIf
% \EndProcedure
% \end{algorithmic}
% \scriptsize
% % Note: For any tree node $s$, $idx_s$ denotes the index of selected node $s$.
% \end{algorithm}
\begin{algorithm}
\caption{An MCTS Iteration on thread $j\in [1,p]$}\label{alg:cpu}
\begin{algorithmic}[1]
\small
% \Procedure{Simulation of a single worker with ST-CPU}{}
% \If {$j==1$ (Master Simulation thread)}
% \State DataMigrate(receiver buffer,$FPGA\rightarrow CPU$); flag$\leftarrow$0;
% \EndIf
\State $s,s'\leftarrow$Selection and Node Insertion using FPGA; 
\State Receive buffer$[j]\leftarrow$ $s,s'$; \Comment{\textcolor{blue}{PCIe O($p$)}}
% \If {flag $==$ 0}
% \State $s,s'\leftarrow$receiver buffer$[j]$;
\State state($s$)$\leftarrow$ST[$s$]; \Comment{\textcolor{blue}{Concurrent read ST}}
\State simulate 1 step with initial state($s$), initial action($s,s'$);
\State ST$[s']\leftarrow$state($s'$)); \Comment{\textcolor{blue}{State Insertion, Concurrent write ST}}
\State $V_j\leftarrow$Simulation phase from state($s'$) until termination;
\State Send buffer$[j]\leftarrow V_j$; 
\State barrier synchronization;
\State migrate $V$ from the Send buffer to FPGA; \Comment{\textcolor{blue}{PCIe O($p$)}}
\State BackUp using FPGA;
% locked(flag$++$);
% % \EndIf
% \If {$j==1$ (Master Simulation thread) and flag$==p$}
%     \State DataMigrate(sender buffer,$FPGA\leftarrow CPU$); 
% \EndIf
% \EndProcedure
% \Procedure{Tree Flush-FPGA}{}
%     \State UCT.root$\leftarrow$ Eq. \ref{eq:uct}; delete non-root nodes and edges; 
%     \State send old index of new root $idx_r$ to CPU;
% \EndProcedure
% \Procedure{Tree Flush-CPU}{}
%     \State receive $idx_r$ from FPGA;
%     \State ST[root index]$\leftarrow$ST[$idx_r$]; empty all non-root node entries; 
% \EndProcedure
\end{algorithmic}
% \scriptsize
% Note: flag is a shared variable for barrier-synchronization between $p$ threads before communication with FPGA.
\end{algorithm}
\subsection{Tree-Parallel MCTS: Bulk Synchronous Processing}
% with Constrained Communication Rule}
\label{sec:bsp}
 The CPU-FPGA system execution model is shown in Fig. \ref{fig:sys}. It can be viewed as a Bulk Synchronous Process (BSP) with constrained communication at the end of each iteration.
%  MCTS also 
% even if multiple workers selected the same $s$, 
% so $p$ parallel insertions into ST (Algorithm \ref{alg:after},line 14) always operate on different indices, without any inter-worker data dependencies.
 
 \textbf{CPU-FPGA Interface and Constrained Communication:}
 As shown in Fig. \ref{fig:sys}, we define a shared Send buffer of size $p$ to store rewards $V$ produced by Simulation threads. Its data is migrated to the FPGA after barrier synchronization of all the threads. After the Selection and Node Insertion on FPGA are completed,  node indices $s,s'$ are written into a Receive buffer where each entry can be independently accessed by a thread. We define a constrained communication rule for the BSP: each thread $j\in [1...p]$ must access the same index $j$ of the Receive buffer and the Send buffer. Similarly, each worker on FPGA also access the same unique index of the Receive buffer and the Send buffer. This ensures the correct matching between $V$ and traversed $E_t$ of each worker in the BackUp phase (Algorithm \ref{alg:in-tree}).
% and receive the nodes $s,s'$ from FPGA, both in a streaming manner. 
% On the FPGA side, we let the accelerator directly transfer data with CPU
% without accessing the FPGA DRAM for low-latency communication.
% to minimize latency. 
% Note that streaming interface ensures the order in which $V$ reaches FPGA is the order in which worker

\textbf{CPU Data Concurrency:} 
% As shown in the right side of Fig. \ref{fig:sys}, 
% After the component decomposition and mapping, only one round-trip of PCIe transfers are required. 
% We define the interface on the CPU Simulation threads to send the rewards $V$ to the FPGA,
The parallel program for each CPU thread is shown in Algorithm \ref{alg:cpu}.
 Note that concurrent accesses to ST during State Insertion do not require any thread-level synchronization, because MCTS guarantee all workers expand different nodes $s'$, and any expanded $s'$ are different from any selected node $s$. This means all the threads write into different ST entries (Algorithm \ref{alg:cpu} line 5), and no ST read (Algorithm \ref{alg:cpu} line 3) depends on the ST write of any other thread. Therefore, ST operations will not become the bottleneck in system throughput when increasing $p$.
 
\textbf{Correctness of System Execution:} 
% \textit{Correctness:} 
Consider a baseline CPU program for $p$-worker in-tree operations (Algorithm \ref{alg:in-tree}). Our CPU-FPGA system yields the exact same outputs as that of a CPU-only system. 
We prove this from two aspects: 
% (1) With the decomposition of the Expansion phase, the CPU-FPGA system still performs the exact same operations as the baseline.
% This is because the operation of adding new nodes to the tree (\ref{alg:in-tree} line 11) is not changed by our decomposition. It is only applied to UCT and ST separately, where UCT and ST are related using the same encoding of node indices.  
% (2) 
The FPGA acceleration avoids any race conditions in BackUp, Selection and Expansion phases.
First, the 3 phases by all workers are performed in sequence to avoid data race between workers in different phases.
Second, the key concept of FPGA acceleration is to utilize a pipeline where each stage only access a particular UCT level. 
The pipeline avoids any RAW (Selection, Expansion) and WAW (BackUp) race conditions between workers by ensuring only one worker is processed in a stage at any time.

% \section{Application Programming Interface}
% 0.5 page
% \vspace{-5pt}
\section{Accelerator Implementation}
\label{sec:arch}
% 1-1.5 page
% \vspace{-5pt}
\subsection{Accelerator Overview}
The overview of the accelerator is depicted in Fig. \ref{fig:arch}. The accelerator is composed of a worker distributor using the root level information to assign workers to different pipelines, and $n$ pipelines working on strictly separate sub-trees. Given the tree fanout $F$, and the number of workers $p$, we allocate pipelines that support concurrency of up to $F$ sub-trees and avoid hardware under-utilization. Thus, we set $n=min(p,F)$.
% PCIe Stream Interface, compute modules for all the in-tree operations, and memory modules managing the UCT graph (memory banks, buffers, crossbar switch, and on-chip FIFO channels). 
As motivated in Sec. \ref{sec:background}, The key objectives of the accelerator design are to 
% (1) achieve low-latency memory accesses and low-latency comparisons over child nodes for the same in-tree worker, and (2) minimize the overhead of synchronization due to RAW dependencies between in-tree workers. 
support low data access latency $T_{mem}$ (Sec. \ref{sec:partitioning}), and minimize both the intra-pipeline $T_{sync}$ (Sec. \ref{sec:quantized}) and  inter-pipeline $T_{sync}$ (Sec. \ref{sec:pipeline}).
\vspace{-5pt}
\subsection{Minimizing $T_{mem.}$}
\label{sec:partitioning}
We adopt a compact adjacency list data structure to store the UCT in on-chip memory banks. 
Each node entry in a memory bank is a node index followed by all of its adjacent edge information (edge weights and child node indices). 
The UCT is partitioned and mapped to the FPGA SRAM. The Root Level Memory Bank stores a single root node entry with fully partitioned adjacency edge array.
% to support concurrent updates. 
To support both data parallelism across pipelines, we partition the UCT into $n$ subtrees rooted at different child node(s) of the root (each stored in a Sub-Tree Memory Bank Group). To support stage parallelism within a pipeline, for each sub-tree, we partition node entries by UCT levels (different levels are stored in separate banks). 
% Each partition is stored in separate Sub-Tree Memory Bank Groups to serve different pipelines. 
% To support pipelining over different tree levels, each memory bank group is partitioned based on tree levels, i.e. 
% and store node entries for different levels in separate banks.
% Since the structure of the tree cannot be determined at compile-time,
At compile time, we statically allocate SRAM banks with the capacities to store a full $F$-ary UCT with height $D$ that can store number of nodes $>X$, where $X$ is the maximum number of tree nodes. 
% We ensure 1-to-1 mapping between SRAM array indices and node indices.
% All in-tree operations directly access SRAM banks at run-time without any hardware re-configuration.
These techniques guarantee no bank conflicts occur between workers,
% in different pipelines and pipeline stages, T
and ensure single-cycle accesses to the UCT by all the workers.
% (compactors, updaters and inserterss).
% is possible. At run-time, the node indices are statically mapped to corresponding memory locations such that reading/writing accesses of pieline stages and pipelines are guaranteed to have zero bank conflict and thus ganrateed single-cycle accesses.
\vspace{-5pt}
\subsection{Minimizing Intra-Pipeline $T_{sync.}$}
\label{sec:quantized}
% \subsubsection{Analysis and Design Space Exploration}
Each stage of the Sub-Tree Selection Pipeline involves a 2-input comparator that loops over $F$ adjacent edges sequentially.
% in its tree level.
% Ideally, a worker assigned to a pipeline waits for the previous worker to finish comparing all edges in exactly $F$ cycles.
% However, 
HLS-generated floating point comparator takes multiple cycles to compute \cite{de2020transformations}, introducing loop-carried dependency overheads.
% latency $C$ that requires $CA$ cycles in comparing $F$ edges. 
To workaround this, we introduce fixed-point representation for edge weights (Eq. \ref{eq:uct}) that supports single-cycle comparisons.
% We introduce a variable-bitwidth fixed-point representation scheme specialized for the UCT graph. 
We assign the integer bit-width of the $uct$ value based on the upper bound of $uct$ (this is obtained by let $V_{\hat{s}}=$ the maximum reward and let $N_s=X,N_{\hat{s}}=1$ in Eq. \ref{eq:uct}),
% which is the sum of the maximum reward and size limit of the tree. We then 
and keep 16 bits for the fractional part of the $uct$ value.
Empirically, the loss in precision for the exploration term in Eq. \ref{eq:uct} is within 0.01\%, which is insignificant compared to typical 1\% to 40\% virtual loss applied to the $uct$ value \cite{treeP,wu-uct}. Therefore the result of Eq. \ref{eq:uct} using fixed-point is the same as using floating point numbers. Comparison of edge weights over $F$ edges can be completed in exactly $F$ cycles using the fixed-point representation.
% \vspace{-10pt}
\subsection{Minimizing Inter-Pipeline $T_{sync.}$}
% the input data are processed in an asynchronous manner
\label{sec:pipeline}
% In-tree modules: the in-tree modules are responsible for performing BackUp, Selection and Node Insertion for $p$ workers upon receiving $p$ rewards from parallel simulation workers, and outputting the leaf nodes and expanded edge to the CPU expansion and simulation workers. 
% % Given $p$ workers and the decision tree with maximum depth $D$ and fanout $F$, BackUp and Selection dominate the computations ($O(p\times D\times A)$ for Selection and $O(p\times D)$ for BackUp) among in-tree operations ($O(p)$ for Node Insertion). Therefore, we exploit both data parallelism and pipeline parallelism for these phases.
% We assume the decision tree size limit (maximum depth $D$ and fanout $F$) and the number of available Simulation workers, $p$ are given at compile time. The dataflow and optimizations for all in-tree operations shown in Fig. \ref{fig:arch} are summarized below.

%  we allocate $n=min(p,A)$ independent data parallel pipelines, each pipeline responsible for traversing, updating and building separate sub-tree(s) branched from separate child node(s) of the root with depth $D-1$.
 
%  The data parallelism $n$ should be bounded by $F$ to achieve maximum root-level parallelism, and bounded by the number of parallel workers to avoid hardware under-utilization. Thus, we set $ n=min(p,A)$.
%  first execute BackUp updates at the root level, then
 The root-level Worker Distributor assigns incoming workers to different Sub-Tree Selection Pipelines based on the maximum of $F$ edge weights at the root. To fully exploit the data parallelism provided by multiple pipelines, the Worker Distributor needs to  process workers with minimized interval (i.e., we need inter-pipeline $T_{sync}<$ intra-pipeline $T_{sync}$). Therefore, we design the Worker Distributor to compute Eq. \ref{eq:uct} in 1 cycle for arbitrary $F$. Specifically, to obtain the maximum of $f$ edge weights, we assign $C_2^f$=$\frac{f!}{(f-2)!2!}$ comparators, each compares a unique pair of edge weights, and outputs a 1-bit result of comparison. The concatenation of these results (a $C_2^f$-bit number) is used to index a Comparison Look-Up Table (CLUT) that outputs the best child index $\hat{s}$ (Eq. \ref{eq:uct}). The CLUT output $\hat{s}$ is connected to a 1-to-$n$ crossbar to distribute the worker to one of $n$ pipelines.
%  for BackUp, Selection and Node Insertion in the same sub-tree bank group. 
 The size of a CLUT is $2^{C_2^f}$ entries to cover all the possible permutations in $f$ edge weights. To constrain the CLUT size for large fanout $F$, we develop multi-level hierarchy of smaller CLUTs. For example, when $F=20$, we distribute the edge weights to 5 CLUTs, each CLUT has $f=4$. The results of these 5 CLUTs in the first level are passed to the next level with a $f=5$ CLUT to output the maximum of 20 edge weights
 and the associated child index. 
 This design allows 1-cycle response to any changes in $F$ edge weights. By consuming additional comparators and memory only in the Root Level (Worker Distributor), we allow concurrent execution of workers to operate on the same UCT level in different pipelines.
 We thus minimized inter-pipeline $T_{sync.}$ to be constantly 2 cycles (1 cycle for comparison, 1 cycle for $VL$ update at the root) when a single CLUT is used.
% TODO: check this sentence. correct?
% Since typical benchmarks of MCTS features sparse reward and long trajectories, it is common in a large early portion of MCTS steps that workers are evenly distributed among root-level branches because the comparison solely depends on the number of times. 
\vspace{-5pt}
\subsection{Other modules and optimizations}
 \textbf{Memoization:} To eliminate the $O(D)$ overhead of sequentially back-tracing from leaf to node in BackUp, a BackUp Memoization Buffer with size of $D-1$ words is associated with each worker to memorize the node entries to be updated in BackUp during Selection. Thus for each worker, BackUp can be completed in 2 cycles.
%  integrated in the Selection pipeline with single-cycle overhead per worker.
 
\textbf{Node Insertion and Tree Flush:} A node insertion is performed by directly writing into the SRAM location of the expanded leaf node returned by the pipeline, after the Selection phase for all the workers are completed. Insertions in different pipelines work concurrently. 
% A counter is associated with each sub-tree memory bank to track the 
% ===========TODO: do we keep this sentence?===========
% In accordance with existing implementations \cite{wu-uct}, we perform complete Tree Flush that only preserves the new root (and its edges) information. 
% We associate a counter with each sub-tree memory bank to track the number of expanded nodes at that level. 
Upon Tree Flush, We clear the counters for tracking expanded child nodes for all node entries, and update the Root Level Memory bank using the new root information. At the same time, the corresponding entries in ST is also deleted on the CPU.
% The in-tree modules in the next MCTS step check the counters to decide which This method avoids expensive hardware reconfiguration or exhaustive erasing of nodes(edges) information in all flushed sub-trees and node-by-node data shuffling between memory banks for different tree levels

\textit{In summary, the hardware optimizations described in this section make the FPGA in-tree accelerator scale to larger number of workers compared with multi-core CPUs by greatly reducing $T_{sync}$ - we are able to achieve $T_{mem}=1$ cycle, inter-pipeline $T_{sync}=2$ cycles for any fanout $F$ that can be contained using a single CLUT, and intra-pipeline $T_{sync}=F+1$ cycles.}

% such that when we increase the number of workers by a factor of $f$, only a trivial additional amortized latency ($(f-1)\times max(A,p)$ cycles*) occurs.}

% \textit{*Proof based on cycle-accurate performance analysis: Overall, assuming the amortized number of workers distributed to a single pipeline is $\frac{p}{n}$, the total number of cycles required to complete the BackUp-Selection-Node Insertion phases for $p$ wrokers is $p+(\frac{p}{n}+D)\times A$. If $p$ is multiplied by a factor of $f$, the additional number of cycles is $\frac{p\times(f-1)\times A}{n}=\frac{p\times(f-1)\times A}{min(A,p)}=(f-1)\times max(A,p)$}.
% % This means that the scalability to larger number of workers $p$ is improved by around a factor of $n$, and the scalability to deeper trees is improved due to the concurrency of $D-1$ stages in a pipeline.}
\vspace{-5pt}
\section{Evaluation}
% (1 $\sim$1.5 page)
The objectives of our work are to achieve low latency in-tree operations in Tree-Parallel MCTS, and improve the scalability of the system throughput to larger number of workers. We first evaluate our optimizations for in-tree operations, and compare them to their baseline performance on a CPU master process  (Sec. \ref{sec:eval_ops}). 
% We then analyze the implications of State Table (ST) on the scalabilty of the throughput of the CPU-FPGA system, and compare the throughput of our system design to a CPU-only baseline (Sec. \ref{sec:eval_sys}).
We then analyze the throughput and scalability of the CPU-FPGA system, and compare with CPU-only baselines (Sec. \ref{sec:eval_sys}).

\subsection{Experimental Setup}
\textbf{Benchmark environments: }
We choose two widely-used game benchmarks for MCTS: the Atari game Pong \cite{wu-uct}, and the board game Gomoku \cite{gomoku}.
For the Atari game, both the Simulation phase and the 1-step simulation for the Expansion phase use OpenAI-gym library. Its action space (i.e., fanout $F$ of the tree) is 6, tree height limit $D$ is 9, and the maximum number of tree nodes $X$ is 56K. 
For Gomoku, in each worker, the Expansion phase expands all $F$ (instead of one) child nodes of each selected node, and the Simulation phase is a DNN inference
% with $F$ output neurons 
that takes a board state as the input and outputs the utilities (expected rewards) for all its child nodes. The DNN is trained with a Gomoku simulation program \cite{gomoku}. The 1-step simulation for Expansion phase also uses the simulation program.
% using DNN-based simulation in MCTS \cite{gomoku} (DNN is trained with public-available program for Gomoku simulation). 
% For the Atari game, we choose Pong with 
% For the Gomoku, 
We set $6$x$6$ board ($F=36$), $D=5$, and use $X=48K$. The choice of $D$ and $X$ in both benchmarks are the same as those in existing CPU implementations. 
% (note: one MCTS step for Gomoku only requires 1333 iterations since each ).

\textbf{Platforms: }
Our CPU baseline experiments are conducted on an Intel(R) Xeon(R) Gold 5120 server with 2 sockets (56 hardware threads in total) at 2.2GHz, and 19MB L3 cache. The CPU-FPGA platform consists of the same CPU and a Xilinx Alveo U200 board \cite{xilinxalveo} connected by PCIe gen4x16. In all the experiments, $p$ denotes both the number of workers and the total number of CPU software Simulation threads. We use one Simulation thread per worker. For Gomoku, $p$ ranges from 2 to 128. For Pong, the OpenAI-gym simulation time for $p<8$ is greater than 99\% of an MCTS iteration time, the advantage of using FPGA when $p<8$ is marginal. So, we show results for larger $p$ that ranges from 8 to 128.

\textbf{Accelerator specifications: }
We develop a parameterized FPGA kernel template using High-Level Synthesis (HLS) for quick customization.
% and easy integration with high-level language and libraries (e.g., Pytorch \cite{pytorch}). 
We follow VITIS development flow \cite{kathail2020xilinx} for bitstream generation. The design achieves 200MHz frequency. OpenCL is used to implement the data transfer between the CPU and FPGA. The resource utilization of our accelerator for both benchmarks obtained at the largest $p$ are shown in Table \ref{tab:res}.

\begin{table}[]
\centering
% \scriptsize
\begin{threeparttable}
\caption{FPGA Resource Utilization}
\label{tab:res}
\begin{tabular}{|l|l|l|l|l|}
\hline
 &
  \textbf{SRAM} &
  \textbf{DSP} &
  \textbf{LUT} &
  \textbf{FF} \\ \hline
\textbf{\begin{tabular}[c]{@{}l@{}}Atari-Pong\\ ($p=128,n=6$)\end{tabular}} &
  \begin{tabular}[c]{@{}l@{}}24 MB\\ (69\%)\end{tabular} &
  \begin{tabular}[c]{@{}l@{}}123\\ (2\%)\end{tabular} &
  \begin{tabular}[c]{@{}l@{}}118K\\ (14\%)\end{tabular} &
  \begin{tabular}[c]{@{}l@{}}160K\\ (9\%)\end{tabular} \\ \hline
\textbf{\begin{tabular}[c]{@{}l@{}}Gomoku\\ ($p=128,n=36$)\end{tabular}} &
  \begin{tabular}[c]{@{}l@{}}16 MB\\ (46\%)\end{tabular} &
  \begin{tabular}[c]{@{}l@{}}465\\ (7\%)\end{tabular} &
  \begin{tabular}[c]{@{}l@{}}308K\\ (36\%)\end{tabular} &
  \begin{tabular}[c]{@{}l@{}}608K\\ (34\%)\end{tabular} \\ \hline
\end{tabular}
    \begin{tablenotes}
      \scriptsize
      \item Note: 
    %   The difference in DSP consumption for the two benchmarks is due to different number of sub-tree pipelines $n$ (Sec. \ref{sec:arch}) and different number of comparators at root-level Worker Distributor. 
    %   For Pong (Gomoku), $n=6 (32).$ 
      For the Worker Distributor, we use a CLUT with $f=6$ for Pong. We use 2 levels of CLUTs, each with $f=6$ for Gomoku.
    %   We allocate enough memory banks that are able to contain complete tree with height $D$ at compile time to support potential asymmetric tree growth.
    \end{tablenotes}
\end{threeparttable}
\end{table}

\subsection{Evaluation of In-Tree Operations}
\label{sec:eval_ops}
\begin{figure}[h]
    \centering
    \includegraphics[width=0.9\linewidth]{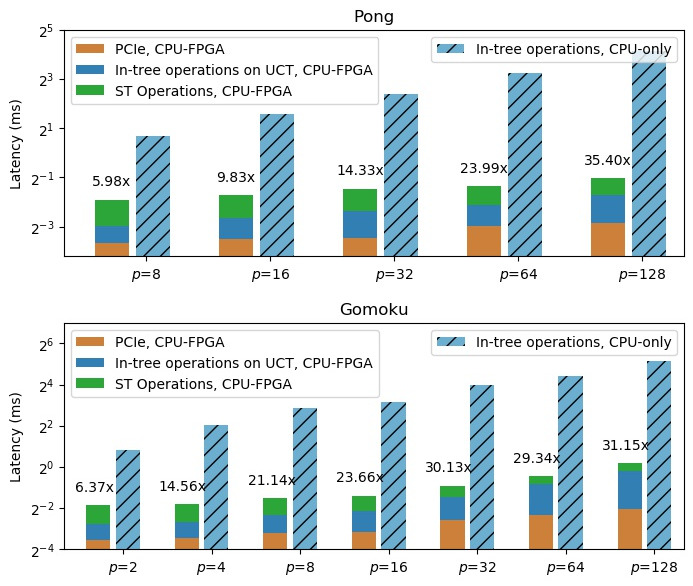}
    \caption{Latency of the in-tree operations. Y-axes are in log scale for better breakdown visualization.}
    \label{fig:latency}
\end{figure}

% Show total latency of Node Insertion-Selection-Backup across 8-64 workers. For FPGA, this should include the PCIe and Simulation Hash latency (show breakdown, suggest system throughput will not be bounded by simulation hash on CPU). Compare with CPU baseline.
% Analyze reasons of speedup (parallelism, pipelining with no stall, single-cycle data accesses)
% \begin{figure}[h]
%     \centering
%     \includegraphics[width=0.98\linewidth]{figs/atari_latency.jpg}
%     \caption{In-tree operations latency for Pong.}
%     \label{fig:atari_latency}
% \end{figure}
Fig. \ref{fig:latency} shows the total latency of in-tree operations in BackUp, Selection and Expansion phases in one MCTS iteration for various number of workers. The reported in-tree operations latency is the average over all the iterations in an MCTS step. For the measurement on our CPU-FPGA system, apart from the FPGA kernel latency, we also include both the PCIe transfer time before Simulation and after BackUp, and the additional ST operations executed on CPU for Expansion. The PCIe data transfer time is obtained using Xilinx Run-time (XRT) Profiler \cite{xrt}.
For a given benchmark, the FPGA kernel performing in-tree operations on the UCT always has lower latency than CPU.
Also, for a given $p$, in-tree operations for Gomoku take longer than that for Pong on CPU, this is because larger $F$ lead to higher $T_{sync}$ in Selection. We point out that the impact of $F$ on Selection latency on the FPGA is smaller, because we optimized $T_{sync}$ between workers to be as low as 2 (3) FPGA cycles for $F=6$ ($F=36$).
% (The hierarchical CLUT for the given $F$ described in Sec. \ref{sec:pipeline} can fit on-chip). 
% (when all workers are distributed to different pipelines at root level). 
As a result, the latency of FPGA kernel for in-tree operations scale well wrt both $F$ and $p$.
The PCIe overhead increases very little as $p$ increases, because the reduced PCIe data transfer only takes negligible amount of time compared to the fixed PCIe initiation latency ($\sim0.04$ ms).
% (e.g. for largest $p=128$, it only transfers )
The tradeoff for lower PCIe data traffic is the additional accesses to ST. 
The ST operations involve reading or writing 256 (432) bytes representation of environment state for Pong (Gomoku).
% The ST operations incur comparable latency to thse PCIe overhead, since the ST accesses involve reading or writing larger amount of data (i.e., complete environment states) than simple rewards and node indices transferred through PCIe. 
The ST operation latency does not incur thread-level synchronization overheads and also scales well to increasing $p$. Overall, we observe a range of $6$ to $35\times$ speedup for in-tree operations. 
% This demonstrates that the in-tree operations in our CPU-FPGA system have better scalability to large $p$ than the CPU-only implementations.

\subsection{Evaluation of System Performance}
\label{sec:eval_sys}
\begin{figure}[h]
    \centering
    \includegraphics[width=\linewidth]{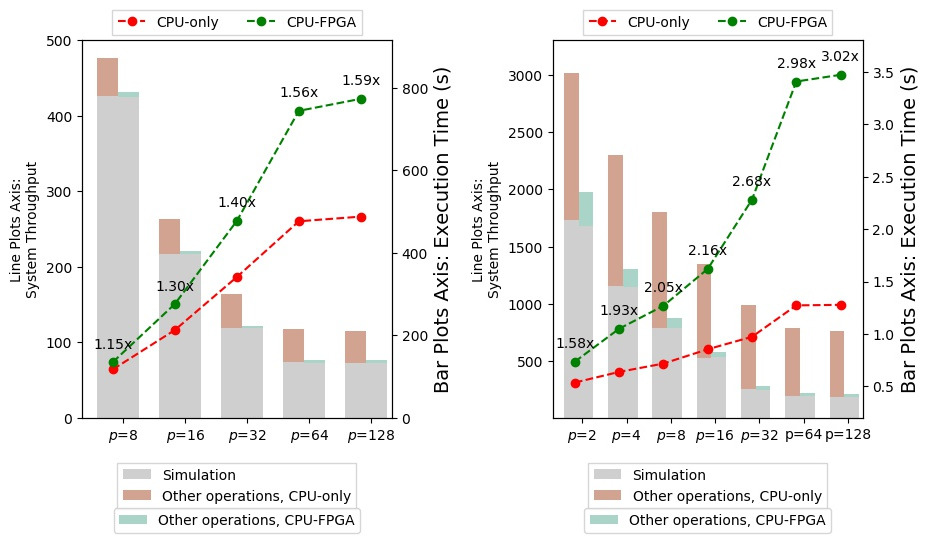}
    \caption{System Throughput for Pong (left) and Gomoku (right). Each throughput marker on a line chart is associated with two background bar plots. These two bars show the complete MCTS step execution time breakdown on CPU-only and CPU-FPGA systems. For the breakdown, ``Simulation" is the total time of the Simulation phase and 1-step simulation for Expansion, it is the same for both CPU-only and CPU-FPGA systems; ``Other operations" includes the total time for in-tree operations in all the MCTS iterations and the Tree Flush time. The $a\times$ text labels above the line chart shows the throughput improvement of CPU-FPGA compared to CPU-only.}
    \label{fig:throughput}
\end{figure}

The \textbf{System Throughput} is defined as the number of simulation requests processed per second. A simulation request refers to one local-environment simulation (or batch-1 DNN inference) by a single worker in an MCTS iteration. Fig. \ref{fig:throughput} shows 
the system throughput of the two benchmarks for various $p$. 
% 2. Algorithm performance with fixed-point data representation for UCT: achieved reward of fixed-point representation vs float-point UCT on CPU is the same.
% Line chart suggests that our acceleration improved throughput and scalability without changing algorithm performance.
% 2. Anatomy and Analysis of Scalability: execution time breakdown (timeline): FPGA kernel, PCIe, Simulation vs CPU in-tree operations, Simulation. FPGA+PCIe has lower end-to-end latency than CPU in-tree operations for larger number of workers. FPGA latency is so optimized such that FPGA+PCIe is largely dominated by PCIe, so that increasing workers does not have large impact on end-to-end latency of each MCTS iteration. 
The CPU-only system throughput improvement cannot linearly scale up with increasing $p$ because the increasing latency of in-tree operations becomes the bottleneck. The proposed FPGA acceleration alleviates this bottleneck by reducing $T_{sync}$ between workers
and improves the system scalability. 
Note that the system throughput stops increasing after $p>56$ because there are only 56 hardware CPU threads.
Note that the speedup using FPGA will further increase if more parallelism were enabled for simulation.
% such that simulation time does not scale down further. 
Higher system throughput improvements from CPU-FPGA system are consistently observed for larger $p$, because when $p$ increases, fewer MCTS iterations are needed to reach $X$ tree nodes in an MCTS step, the total simulation time decreases, while the total in-tree operations time on CPU does not drop significantly such that its bottleneck becomes more significant. 
For the Gomoku benchmark, for large $p$, the gap between ideal speedup and actual throughput on the CPU is especially big because the latency of largely sequential in-tree operations exceeds that of the Simulation phase (parallel DNN inference). 
Our system closed such gap by realizing low-latency in-tree operations, thus improved the scalability to larger $p$ in both benchmarks. 
Overall, we obtain up to $1.59\times$ and $3.02\times$ higher throughput for the two benchmarks than the CPU-only baselines.

% Note the PCIe overhead is non-negligible compared with in-tree operations, but still more scalable than using CPU implementation since we scale better to more workers with a near-constant PCIe overhead. Link to Discussion on future work to support general DNN inference-based simulation on FPGA to eliminate PCIe overhead in Sec. \ref{sec:conclusion}.
\vspace{-5pt}
\section{Related Work}
\subsection{MCTS Parallelization}
\label{sec:related_parallel}
In addition to Tree-Parallel MCTS, other algorithms have been developed to parallelize MCTS. The objective is to speedup each MCTS step while reducing the negative impact on algorithm performance (algorithm performance is measured by achievable reward using a fixed total number of simulations). Popular parallelization approaches include Leaf-Parallel MCTS (LeafP) \cite{leafP} that parallelizes simulations at the same leaf node,
% which affect the algorithm,
and
Root-Parallel MCTS (RootP) \cite{rootP} that create multiple trees at different workers and aggregates their statistics of the subtrees before all the workers complete an MCTS step.
% their tree constructions.
\cite{rocki2011parallel,barriga2014parallel} proposed block parallelism - a combination of  leafP and rootP for GPU acceleration.
% In this work, 
% we target Tree-Parallel MCTS (TreeP).
% due to its ability to maintain superior algorithm performance compared with the other parallelization approaches, which is shown by recent work that proposed WU-UCT, a variant of Tree-parallel MCTS\cite{wu-uct}. 
Recent work WU-UCT \cite{wu-uct} proposes a variant of Tree-Parallel MCTS (TreeP), and shows that it maintains superior algorithm performance compared with the other parallelization approaches. 
In this work, we target TreeP because of its superior algorithm performance.
The difference in variants of Tree-Parallel MCTS is in how the virtual loss \textit{VL} is computed: it can be a value proportional to the number of visits to child nodes as proposed in \cite{wu-uct}, or a constant \cite{treeP}. Note that our accelerator design can support both these variants  by a simple modification to the Sub-Tree Selection pipelines.
% Tree paralellization avoids the collapse of exploration caused by multiple workers landing on the same leaf node during Selection. 
% Recent work \cite{wu-uct} have shown that to achieve a certain reward, Tree-parallel MCTS by applying a simple virtual loss of pre-updating $N_s$ and $N_s'$ during the multi-worker selection phase requires almost the same total number of simulations compared to sequential MCTS.s
% \vspace{-15pt}
\subsection{Hardware-Accelerated MCTS}
\label{sec:related_hardware}
There is limited work in hardware acceleration of MCTS. 
\cite{jahanshahi2013blokus,qasemi2014highly} design Blokus Duo Game solvers on FPGA that uses MCTS. Their accelerators target Blokus Duo game only and implement the simulator circuit on FPGA. 
Different from their work, Our objective is to develop a general system for MCTS by keeping software simulations on the CPU.
% Thus, \cite{jahanshahi2013blokus,qasemi2014highly} target a different scope and are not directly comparable to our work.
% and only focus on in-tree operations. 
% \cite{rocki2011parallel,barriga2014parallel} accelerate parallel simulations in MCTS using a combination of RootP and LeafP and map them onto GPU thread blocks. Since the GPU parallelization approaches used are different than the TreeP targeted in our work, they do not have the challenges of thread-level synchronizations and are not  fairly comparable to our work.

% Summarize the two, and why we do not compare with them in eval.

\section{Conclusion \& Future Work}

\label{sec:conclusion} 
In this work, we proposed the first CPU-FPGA system design for accelerating Tree-Parallel MCTS that is portable to various applications. 
% We applied novel optimizations in both system-level task organization and hardware acceleration. 
Our system achieved higher throughput and better scalability to larger number of parallel workers than state-of-the-art CPU-only implementations.

Parallel MCTS also lead to many further opportunities for efficient acceleration. For example, 
% by developing an overlay architecture for DNN-based MCTS, where the DNN is implemented on the FPGA and ST is managed using FPGA DRAM resources, the need for repeated PCIe transfers within a MCTS step can be eliminated.
our accelerator design poses a limit on the supported tree height since it allocates SRAM banks for the full tree. In the future, runtime dynamic SRAM bank management methodologies can be explored to better support arbitrary-shaped tree growth.
Additionally, new algorithmic optimizations can be explored to reduce data race between workers in TreeP and enable higher data parallelism.

\section{Acknowledgement}
This work is supported by NSF under grant No. CNS-2009057.

\newpage
\clearpage
\bibliographystyle{IEEEtran}
% \IEEEtriggeratref{18}
\bibliography{ref}

\end{document}